\documentclass[aps,prd,amsmath,twocolumn,superscriptaddress,floatfix]{revtex4}
\usepackage[dvips]{color}

\newcommand{\reff}{1}
\newcommand{\hir}{H_{1,i}}
\newcommand{\ec}{\epsilon_c}

	\usepackage{amsmath}
	\usepackage{amsfonts}
  	\usepackage{amssymb}
  	\usepackage{float}
	\usepackage{makeidx}
	\usepackage{amsfonts}
	\usepackage[ansinew]{}
	\usepackage[usenames,dvipsnames]{pstricks}
	\usepackage{epsfig}

\usepackage{xcolor}

\newcommand{\be}{\begin{equation}}
\newcommand{\ee}{\end{equation}}
\newcommand{\bea}{\begin{eqnarray}}
\newcommand{\eea}{\end{eqnarray}}
\newcommand{\eqn}[1]{(\ref{#1})}

\newcommand{\eq}[1]{Eq.~(\ref{#1})}
\newcommand{\eqs}[1]{Eqs.~(\ref{#1})} 
\newcommand{\Mp}{M_{\rm Pl}}
\DeclareMathOperator{\sign}{sign}

\makeindex

\begin{document}


\title{Reconstructing homospectral inflationary potentials}
\author{Alexander Gallego Cadavid}
\affiliation{Universidad  de Valpara\'{\i}so,  Avenida Gran Breta\~na 1111,  Valpara\'{\i}so  2360102,  Chile}
\affiliation{Instituto de F\'isica, Universidad de Antioquia, A.A.1226, Medell\'in, Colombia}
\author{Antonio Enea Romano}
\affiliation{Instituto de F\'isica, Universidad de Antioquia, A.A.1226, Medell\'in, Colombia}
\affiliation{ICRANet, Piazza della Repubblica 10, I--65122 Pescara, Italy}
\author{Andrew R. Liddle}
\affiliation{Instituto de Astrof\'{\i}sica e Ci\^{e}ncias do Espa\c{c}o,
Faculdade de Ci\^{e}ncias, Universidade de Lisboa, 1769-016 Lisboa, Portugal}

\date{\today}

\begin{abstract}
Purely geometrical arguments show that there exist classes of homospectral inflationary cosmologies, i.e.\ different expansion histories  producing the same spectrum of comoving curvature perturbations. We develop a general algorithm to reconstruct the potential of minimally-coupled single scalar fields from an arbitrary expansion history. We apply it to homospectral expansion histories to obtain the corresponding potentials, providing numerical and analytical examples.
The infinite class of homospectral potentials  depends on two free parameters, the initial energy scale and the initial value of the field, showing that in general it is impossible to reconstruct a unique potential from the curvature spectrum unless the initial energy scale and the field value are fixed, for instance through observation of primordial gravitational waves.
\end{abstract}

\pacs{Valid PACS appear here}
\maketitle


\section{Introduction}

Using purely geometrical arguments, it has been shown \cite{GallegoCadavid:2017wng} that there exist  infinite classes of homospectral inflationary cosmologies, whereby different expansion histories produce the same spectrum of comoving curvature perturbations. 
This is because the equation for the evolution of comoving curvature perturbations is completely determined by a function $z(t)$ of the scale factor $a(t)$, involving first- and second-order time derivatives. This implies an infinite class of expansion histories giving the same $z(t)$, corresponding to different initial values of the first derivative of the scale factor, or in physical terms, to the initial energy scale.

In this paper we aim to bridge the gap between geometry and physics by developing a general algorithm to reconstruct the potential of minimally-coupled single scalar fields from any arbitrary expansion history, which we can then apply to generate homospectral expansion histories.
The infinite class of homospectral potentials obtained in this way, each corresponding to a different initial energy scale, defines  models with the same spectrum of curvature perturbations, showing that in general it is impossible to reconstruct a unique potential from the curvature spectrum, unless the initial energy scale is fixed (for instance through observation of primordial tensor perturbations).
We show that for any spectrum of comoving curvature perturbations there exists an infinite class of homospectral single-field models, and explicitly compute the corresponding potentials.

As an example we explicitly reconstruct numerically some homospectral potentials from their corresponding expansion histories.
Our arguments are model independent and can be generalized to other more complicated theoretical scenarios for which the Sasaki--Mukhanov equation \cite{Mukhanov:1988jd, MUKHANOV1992203, 10.1143/PTP.76.1036} applies, since the origin of the existence of these homospectral models is geometrical, and the evolution of comoving curvature perturbations is completely determined by the expansion history.

\section{Construction of homospectral models}
The construction of homospectral models is based on the fact that the primordial spectrum of curvature perturbations is completely determined by a function $z(t)$ defined below, but that there is an infinite set of scale factor functions $a(t)$ corresponding to the same $z(t)$, due to the freedom in the choice of the initial value of the Hubble parameter $H_i$. From now on a sub-index $i$ denotes values of quantities evaluated at the initial time.

The equation for curvature perturbations on comoving slices $\mathcal{R}_{c}$ is \cite{Mukhanov:1988jd, MUKHANOV1992203, Garriga:1999vw,10.1143/PTP.76.1036, Langlois:2010xc} 
\begin{equation}\label{eq:cpe}
  \mathcal{R}_{c}''(k) + 2 \frac{z'}{z} \mathcal{R}_{c}'(k) + c^2_{\rm s} k^2 \mathcal{R}_{c}(k) = 0,
\end{equation}
where $k$ is the comoving wave number, $c_{\rm s}$ is the sound speed, and primes indicate derivatives with respect to conformal time $d\tau \equiv dt/a$. It can be seen that the functions $z$ and $c_{\rm s}$ completely determine the spectrum. They can be written in terms of the scale factor $a(t)$ as \cite{GallegoCadavid:2017wng}
\begin{equation}\label{eq:z}
c_{\rm s} z \equiv 
a\sqrt{2\epsilon}= \sqrt{2\Biggl(a^2 - \frac{ a^3 \ddot{a}}{\dot{a}^2}\Biggl)},
\end{equation}
where dots indicate derivatives with respect to cosmic time, 
\be
\epsilon \equiv -\frac{\dot H}{H^2},
\ee
is the slow-roll parameter, and $H\equiv\dot{a}/a$ is the Hubble parameter.

For a given function $z_\reff$ the scale factor evolution is not uniquely determined. From the above relation we get in fact a second-order differential equation for the scale factor
\begin{equation}\label{eq:diffa}
a^2 - \frac{ a^3 \ddot{a}}{\dot{a}^2} = \frac{1}{2} z_\reff^2 c_{{\rm s},\reff}^2= a_\reff^2 \epsilon_\reff \,.
\end{equation}
The initial value of the scale factor has no physical importance since it can always be arbitrarily rescaled, but the initial condition for the first time derivative is physically important since it corresponds to considering background histories with different \textit{initial Hubble} parameters $H_i$, and hence different \textit{initial energy scales}. 

We will parametrize this difference in the initial energy scale with the dimensionless quantity $H_{2,i}/H_{1,i}$, where the subscripts stand for two different members of the same homospectral class.
 This freedom in choosing the initial value of $H_i$ while keeping the same evolution of the function $z(t)$ is the origin of the existence of an infinite set of expansion histories producing the same spectrum of curvature perturbations, which was found in some specific classes of models in Refs.~\cite{GallegoCadavid:2017bzb,GallegoCadavid:2017wng, GallegoCadavid:2017pol}.
 
\section{Reconstruction of the potential from the expansion history}
\label{sec:3}

The Friedmann and acceleration equations for the inflaton field $\phi$ are given respectively by
\bea
 \left(\frac{\dot a}{a}\right)^2 &=& \frac{1}{3 M^2_{\rm Pl}}\left[ V(t) + \frac{1}{2} \dot \phi^2 \right], \label{fei} \\ 
  \frac{\ddot{a} }{a} &=& \frac{1}{3 M^2_{\rm Pl}} \left[ V(t) - \dot \phi^2  \right] \label{aei} \,,
\eea
where $M_{\rm Pl} \equiv (8 \pi G)^{-1/2}$ is the reduced Planck mass and $V$ is the potential energy. Combining these 
we obtain the useful relation
\be
V(t)=M_{\rm Pl}^2\left(3 H^2+\dot{H}\right)=M_{\rm Pl}^2 \left[2\left( \frac{\dot{a}}{a}\right)^2+\frac{\ddot{a}}{a}\right] ,
\label{Vt}  
\ee
which relates directly the physics, i.e.\ the potential $V(t)$, to the geometry, i.e. the scale factor $a(t)$, and allows us to reconstruct the potential producing a given expansion history. As a final step to get the potential as a function of the field it is necessary to determine the function $\phi(t)$ and invert it to get $t(\phi)$.

Summarizing, these are the steps for obtaining the potential $V(\phi)$ from a given expansion history $a(t)$:
\begin{itemize}
    \item Compute $V(t)$ from $a(t)$ using  Eq.~(\ref{Vt}).
    \item Substitute $V(t)$ in one of the Einstein's equations and solve for $\phi(t)$.
    \item Invert $\phi(t)$ to get $V(\phi)=V(t(\phi))$. 
\end{itemize}
Note that the last step gives a unique potential only if  there is a one-to-one correspondence between $\phi$ and $t$, so that for example for an oscillatory $\phi(t)$ the reconstructed potential could be non-unique.
\begin{figure*}
    \centering
    \includegraphics[scale=1.3]{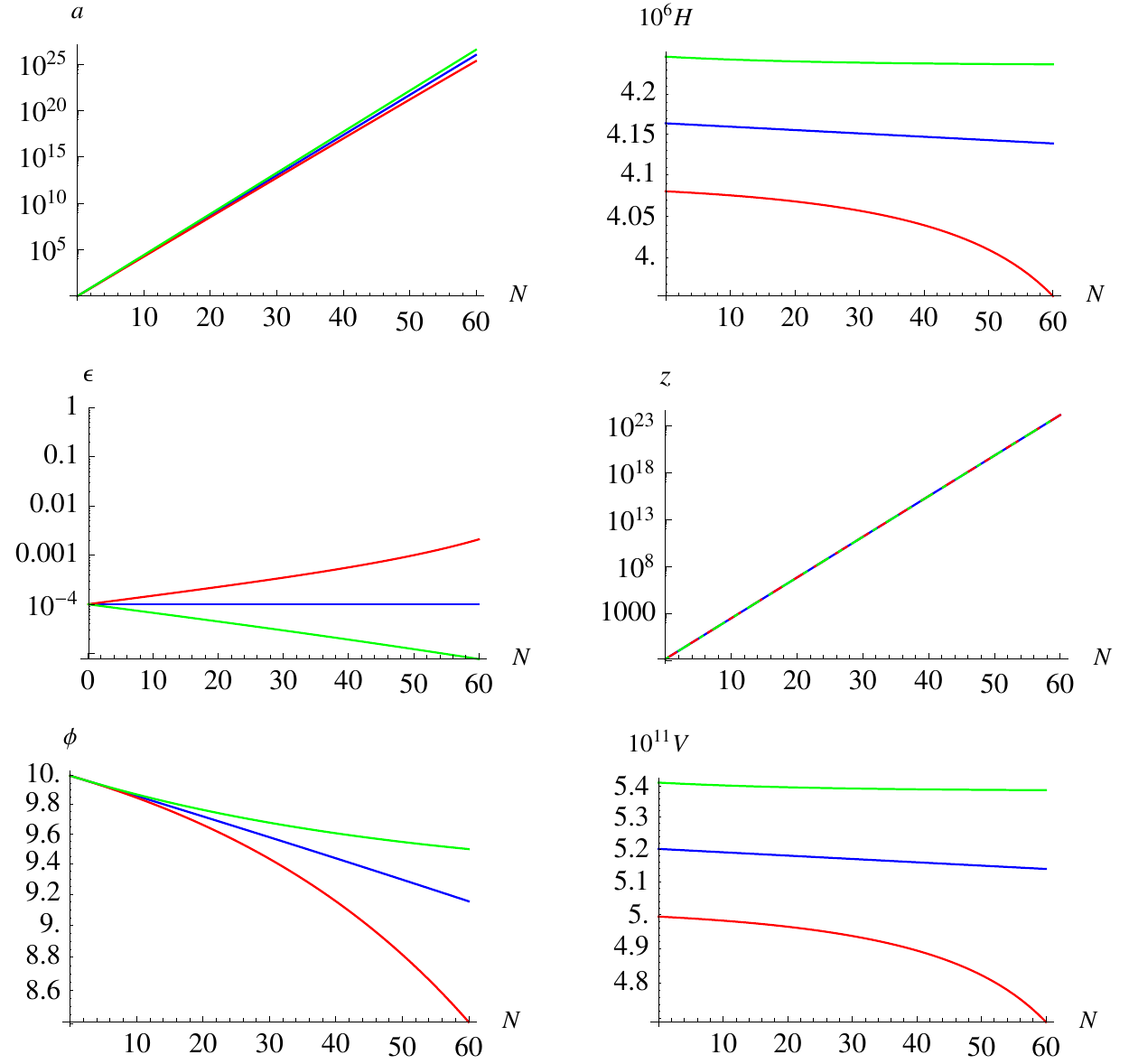}
    \caption{As a function of $e$-folds $N$ for different homospectral models, we show (from top left to bottom right) the scale factor $a$, the Hubble parameter $H$, the slow-roll parameter $\epsilon$, the function $z$ whose lines by construction coincide, the scalar field $\phi$, and its potential $V$ (dimensionful quantities are shown in reduced Planck units). In the plots we use $H_{2,i}/H_{1,i}=1,0.98, 1.02$ for the blue, red, and green lines, respectively.}
    \label{fig:my_label} \label{fig:a}
\end{figure*}

\begin{figure}
    \includegraphics[scale=0.6]{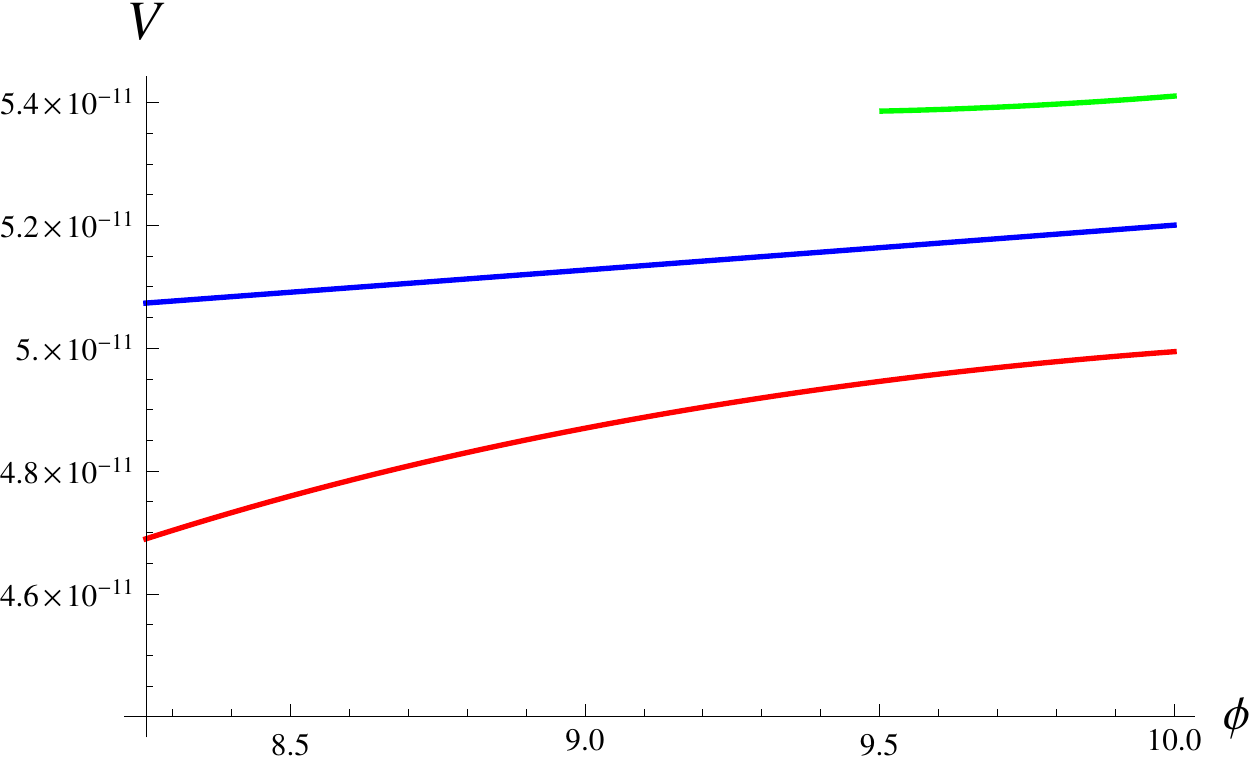}
    \caption{As Fig.~\ref{fig:a}, showing the potential as a function of the scalar field. The field range for different lines correspond to the same
$e$-folds interval plotted in the final panel of Fig.~\ref{fig:a}.}
    \label{fig:Vphi}
\end{figure}

\begin{figure}
    \includegraphics[scale=0.85]{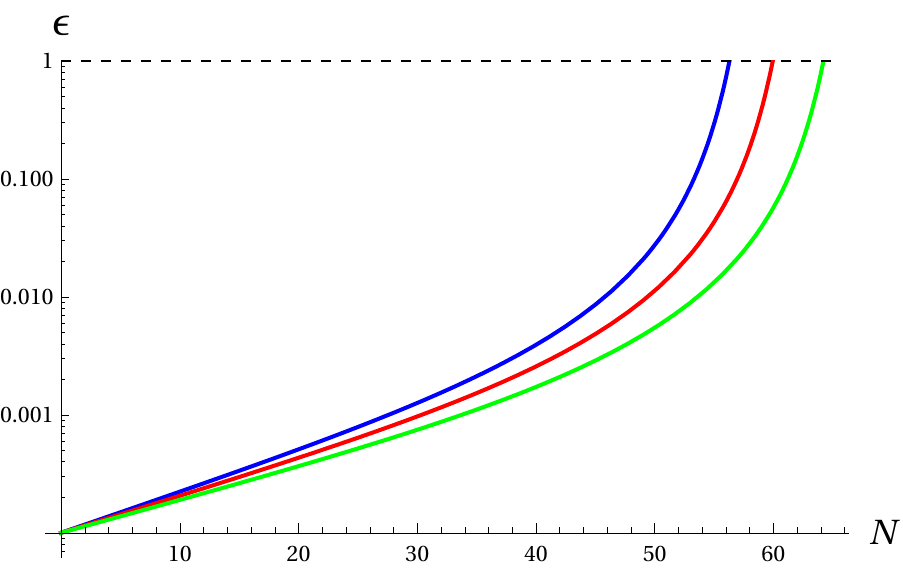}
    \caption{The slow-roll parameter $\epsilon$ is plotted as a function of $e$-folds $N$ showing models amongst the class where inflation ends through $\epsilon=1$. In the plots we use $H_{2,i}/H_{1,i}=0.960,0.964, 0.968$ for the blue, red, and green lines, respectively.}
    \label{fig:epsreachingone}
\end{figure}

\section{Homospectral potentials}

The procedure to obtain a homospectral potential from a given expansion history $a_1(t)$ can be summarized as follows:
\begin{itemize}
    \item compute a homospectral expansion history $a_2(t)$ from $a_1(t)$ using the procedure developed in Ref.~\cite{GallegoCadavid:2017wng}, based on choosing a different initial value $H_{2,i}$ of the Hubble parameter. 
    \item compute $V_2(\phi)$ from $a_2(t)$ following the algorithm outlined in Section \ref{sec:3}.
\end{itemize}

In order to show a concrete example, we will choose a scale factor given by~\cite{GallegoCadavid:2017wng}
\be \label{eq:aref1}
a_\reff (t)=a_{1,i} \left[ 1+ \epsilon_{\mbox{\tiny c}} H_{1,i} (t-t_i) \right]^{1/\ec} ,
\ee
which corresponds to  $\epsilon_\reff(t) = \epsilon_c$, with $\epsilon_c$ constant and $c_{\rm s}=1$. Although we are choosing a specific form for $a_\reff(t)$ the reader should keep in mind that our treatment is completely model independent, and could be applied to any function $a(t)$. Moreover, notice that in the limit $\ec \to 0$, 
\be\label{limita}
a_{\reff}(t) \to a_{1,i} e^{ H_{1,i} (t-t_i) },
\ee
which will be useful to find analytic solutions for the homospectral models as shown in the next section. 

We show the results for the numerical reconstruction of the homospectral potentials in Figs.~\ref{fig:a} and \ref{fig:Vphi}, where we use as time variable $N \equiv \ln (a/a_i)$,  the number of $e$-folds since the beginning of the computation. The values of the parameters used in the figures are $\epsilon_c=10^{-4}$, $c_{\rm s}=1$, $t_i=0$, $a_{1,i}= 1$, and $\phi_{1,i}= 10 \Mp$. 

Note that for a given initial $H_i$, $\phi_2(t_i)$ could be different from $\phi_1(t_i)$, since the choice of the initial value of field is an arbitrary initial condition for the Einstein's equations. This freedom corresponds to obtaining the same $a(t)$ by slow-rolling down different potentials in different field ranges, and implies that while the homospectral expansion histories are a \textit{one-parameter} class, the homospectral potentials are a \textit{two-parameter} class, i.e.\ even for the same value of $H_i$, there can be an infinite class of different potentials giving rise to the same curvature perturbation spectrum. We will explore this extra degeneracy in a separate work, while in this paper we will set $\phi_2(t_i)=\phi_1(t_i)$. Examples of different homospectral potentials  are shown in Fig.~\ref{fig:Vphi}.

In order to check that the homospectral potentials indeed correspond to the same spectra, we substitute the different reconstructed potentials into Eqs.~(\ref{fei}) and (\ref{aei}), solve for $a(t)$ and $\phi(t)$, and compute $z(t)$. The plots of $z$ show that all these models do indeed have the same spectra, since they correspond to the same functional behavior of $z(t)$ \cite{GallegoCadavid:2017wng}.

To align the spectra to the same cosmological scales, it is necessary that inflation ends after the appropriate number of $e$-foldings. This typically requires some other mechanism to intervene, such as a hybrid-inflation style phase transition. 
There will however usually be models amongst the class where inflation ends through $\epsilon$ reaching unity after an appropriate time; see Fig.\ \ref{fig:epsreachingone} for examples. In a sense, those models would be more compelling accounts of the observed spectrum, as they do not need anything extra to self-consistently end inflation.

\section{Analytic approximations for the homospectral models}\label{aahm}

We have shown in the previous section some examples of homospectral potentials reconstructed numerically from different expansion histories, while in this section we will derive some approximate analytical solutions of the reconstruction equations, to provide further insight.
Let us begin by finding an analytic expression for $V_\reff(\phi)$ using \eq{eq:aref1}, following the same procedure outlined previously. First we substitute $a_\reff(t)$ into \eq{Vt} obtaining 
\be
V_\reff (t)= \frac{\Mp^2 \hir^2}{( 1+ \epsilon_c \hir \, t )^2} (3- \epsilon_c )  ,
\ee
which substituted in one of the Einstein's equations gives the following solution for  $\phi_\reff(t)$ 
\be \label{phitref}
\phi_{\reff}(t)= \phi_{1,i} - \sqrt{\frac{2}{\epsilon_c}} \Mp \ln \Bigl[1+ \epsilon_c \hir \, t \Bigr].
\ee

Finally we invert the previous equation to obtain $t=t(\phi_\reff)$, and then replace it in $V_\reff(t)$ yielding
\be\label{Vref}
V_\reff(\phi)= \Mp^2 \hir^2(3- \epsilon_c ) \, e^{\sqrt{2\epsilon_c} \left(\phi-\phi_{1,i}\right)/\Mp}  \, .
\ee

To obtain an analytic approximation for the scale factor of the homospectral models $a_2$, it is convenient to re-write \eq{eq:diffa} as
\begin{equation}\label{eq:diffa2}
\ddot{a}_2= \left(\frac{ \dot{a}_2 }{a_2} \right)^2 \left( 1 - f(t) \right) a_2\,,
\end{equation}
where
\be
f(t) \equiv \left(\frac{a_\reff}{a_2}\right)^2 \epsilon_\reff \,.
\ee

\eq{eq:diffa2} has a solution of the form
\be
a_2(t)= C_2 \exp\left[ \int_1^t \frac{1}{\int_1^x (C_1 + f(y) ) dy} dx\right],
\ee
where $C_1$ and $C_2$ are constants of integration. In the case of \eq{eq:aref1},
a valid approximation for $f(t)$ under  slow-roll approximation is given by 
\be \label{f}
f(t) \approx \left[ \frac{e^{ \hir (t-t_i)} }{e^{H_{2,i} (t-t_i)} } \right]^2 \ec= e^{-2 \hir(H_{2,i}/H_{1,i}-1) (t-t_i)} \ec.
\ee
Using this approximation, with $t_i=0$ and $a_{2,i}=a_{1,i}$, we find this solution for \eq{eq:diffa2} 
\be\label{aa}
a_2(t)= a_{1,i} \left[ \frac{1}{\alpha -1}\left(\alpha \, e^{2 \hir (H_{2,i}/H_{1,i}-1) t} -1 \right) \right]^{1/\alpha \ec} ,
\ee
where
\be
\alpha \equiv \frac{2(H_{2,i}/H_{1,i}-1)}{\ec   (H_{2,i}/H_{1,i})} +1 ,
\ee
with $H_{2,i}/H_{1,i} \ne 1$. In the limit $\ec \to 0$ the analytic solution in \eq{aa} can be written as
\be
a_2(t)=a_{1,i} e^{ H_{2,i} t } ,
\ee
in agreement with \eq{limita}. 

Using the approximation for the scale factor in \eq{aa}, the slow-roll parameter $\epsilon_2$ is given by
\be\label{epsa}
\epsilon_2(t)= \epsilon_c \, e^{-2 \hir (H_{2,i}/H_{1,i}-1) t} .
\ee

\begin{figure}
\begin{minipage}{.45\textwidth}
    \includegraphics[scale=0.6]{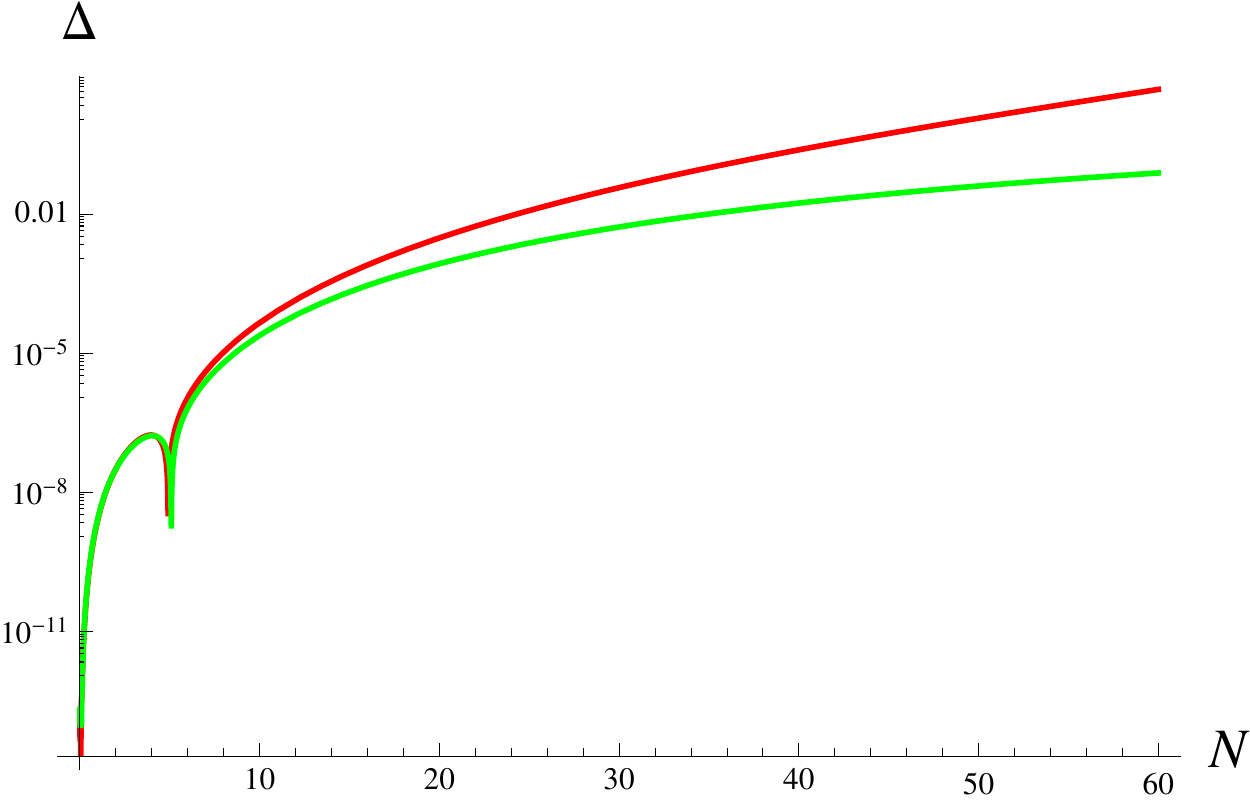}
    \end{minipage}
\begin{minipage}{.45\textwidth}
    \includegraphics[scale=0.6]{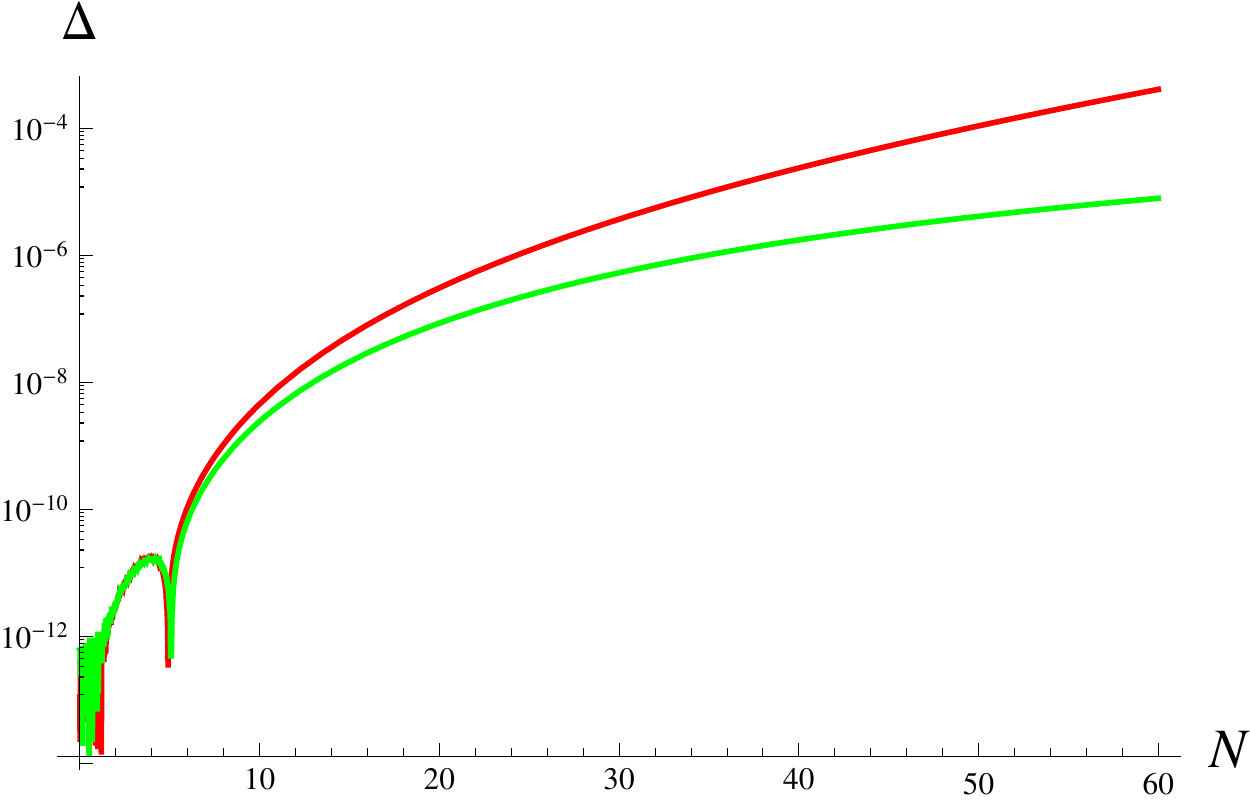}
    \end{minipage}
    \caption{The percentage difference $\Delta \equiv |(a_{\rm num}-a_{\rm ana})/a_{\rm num}|\times 100\%$ is plotted for different models of the same homospectral class as a function of the $e$-folds number. In the plots we use $H_{2,i}/H_{1,i}=0.98 \, (1.02)$ for the red (green) lines and $\ec=10^{-4} \, (10^{-6})$ in the top (bottom) panel.}
    \label{fig:apd}
\end{figure}

\begin{figure}
\begin{minipage}{.45\textwidth}
    \includegraphics[scale=0.6]{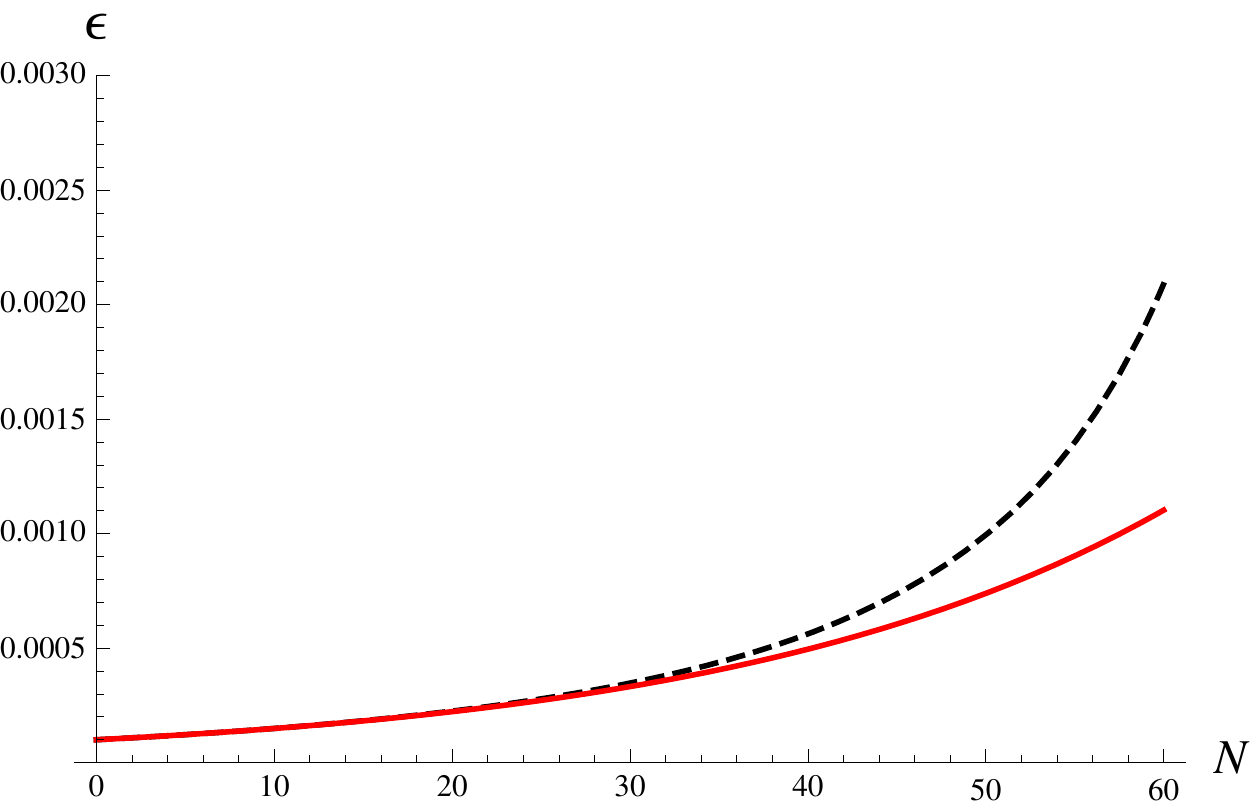}
    \end{minipage}
    \begin{minipage}{.45\textwidth}
    \includegraphics[scale=0.6]{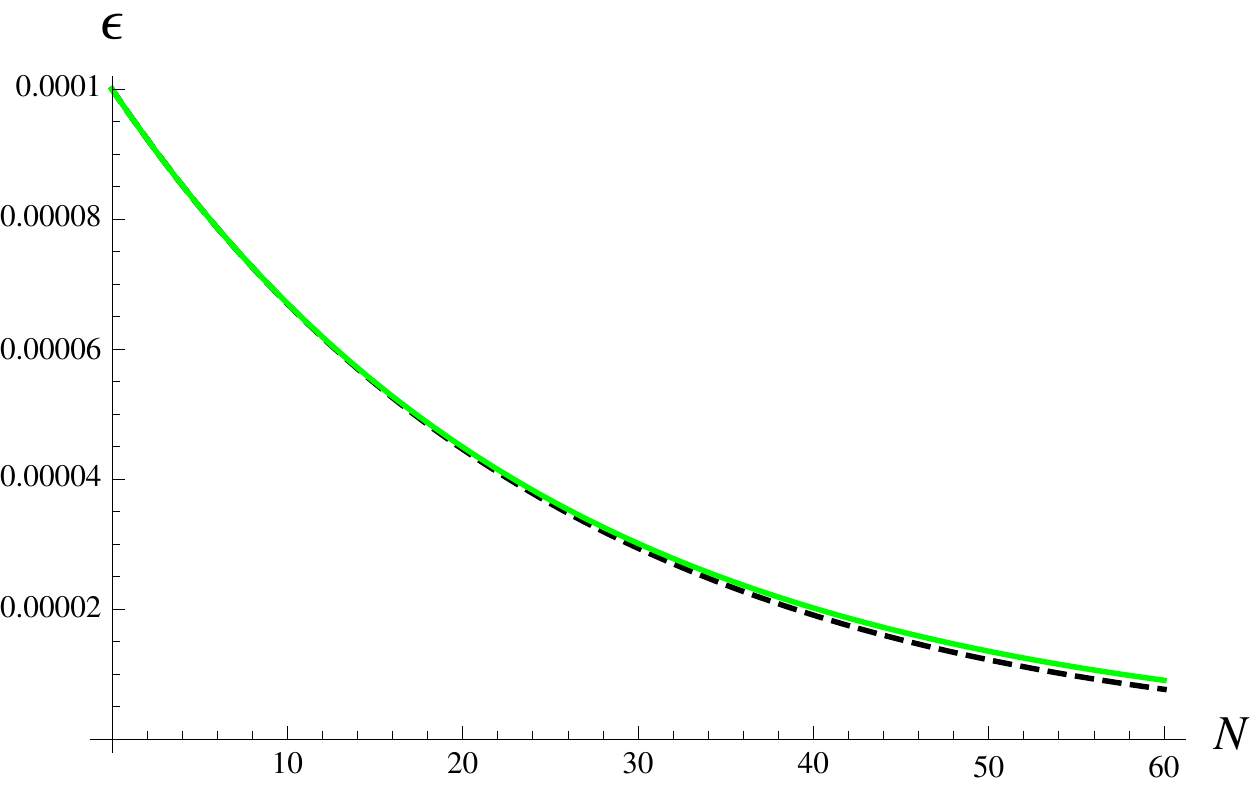}
    \end{minipage}
    \caption{The numerical (black) and analytic (red and green) slow-roll parameters $\epsilon$ are plotted for the same homospectral class. In the plots we use  $H_{2,i}/H_{1,i}= 0.98 \,(1.02)$ in the top (bottom) panel  and $\ec=10^{-4}$.}
    \label{fig:epsplota}
\end{figure}

In Fig.~\ref{fig:apd} we compare the numerical solution of \eq{eq:diffa2} and the corresponding analytic solution in \eq{aa} with $a_{1,i}= 1$. As it can be seen the approximation is in good agreement with the numerical results. 
In Fig.~\ref{fig:epsplota} we compare the slow-roll parameter $\epsilon_2$ using the numerical solution and the approximation in \eq{epsa}. 

In the case of power-law inflation \cite{Lucchin:1984yf}, for which $a_\reff = (t/t_i)^p$ with $p>1$, the approximation 
\be
f(t) \sim \Bigl( \frac{t}{t_i} \Bigr)^\beta,
\ee
is also accurate. Nonetheless we use $a(t)$ given in \eq{eq:aref1} since in the power-law inflation case the analytical treatment is more cumbersome.

Now that we have an analytical approximation for the homospectral scale factors $a_2(t)$, we can compute analytically the corresponding  potentials $V_2(\phi)$, using the same algorithm used for the numerical reconstruction. First we substitute $a_2(t)$ given by \eq{aa} into \eq{Vt} obtaining 
\bea
V_2(t)\!\!&=\!\!& \left[ \frac{2 \hir(H_{2,i}/H_{1,i} -1) e^{2 \hir(H_{2,i}/H_{1,i} -1) t} \Mp }{\alpha e^{2 \hir(H_{2,i}/H_{1,i} -1) t} -1}\right]^2  \notag \\ 
&& \times \left[3 - \epsilon_c e^{-2 \hir(H_{2,i}/H_{1,i} -1) t} \right] .
\eea
Second we substitute $V_2(t)$ into one of the Einstein equations and solve the differential equation for $\phi_2(t)$ with $\phi_2(t_i)=\phi_\reff(t_i)$. In this case we were able to find an analytic expression for  $\phi_2(t)$ given by
\bea \label{phit}
\phi_2(t) &=& \phi_{2,i} - \sign (H_{2,i}/H_{1,i} -1) \Mp \sqrt{\frac{2}{\epsilon_c \alpha}} \\
&& \times \ln \left[ \frac{(\sqrt{\alpha} e^{\hir(H_{2,i}/H_{1,i} -1) t} -1 )(\sqrt{\alpha} +1 )}{(\sqrt{\alpha} e^{\hir(H_{2,i}/H_{1,i} -1) t} +1 )(\sqrt{\alpha} -1 )} \right] . \notag \eea
Finally we invert $\phi_2(t)$ and replace $t(\phi_2)$ in $V_2(t)$ obtaining
\bea \label{Va}
V_2(\phi)&=& \left( \frac{\Mp \, H_{2,i}}{4 \alpha} \right)^2 e^{-2\Phi}  \notag \\
&& \biggl( (\sqrt{\alpha} +1) e^{\Phi} + (\sqrt{\alpha} -1) \biggr)^2 \notag \\
&&  \biggl[ 3 \biggl( (\sqrt{\alpha} +1) e^{\Phi} + (\sqrt{\alpha} -1) \biggr)^2  \\
&&- \alpha \epsilon_c \biggl( (\sqrt{\alpha} +1) e^{\Phi} - (\sqrt{\alpha} -1) \biggr)^2
\biggr] \notag ,
\eea
where
\be
\Phi \equiv \sqrt{\frac{\alpha \epsilon_c}{2}} \frac{(\phi-\phi_{2,i})}{\Mp} .
\ee

The only approximation made to obtain the potential in \eq{Va} is the one in \eq{f}. Notice also that, although $\alpha$ can be negative for \mbox{$H_{2,i}/H_{1,i}<1$,} all previous results are real-valued. As a consistency check it easy to verify that in the limit $H_{2,i}/H_{1,i} \to 1$, \eqs{phit} and \eqn{Va} coincide with \eqs{phitref} and \eqn{Vref}, respectively.

\begin{figure}
\begin{minipage}{.45\textwidth}
    \includegraphics[scale=0.6]{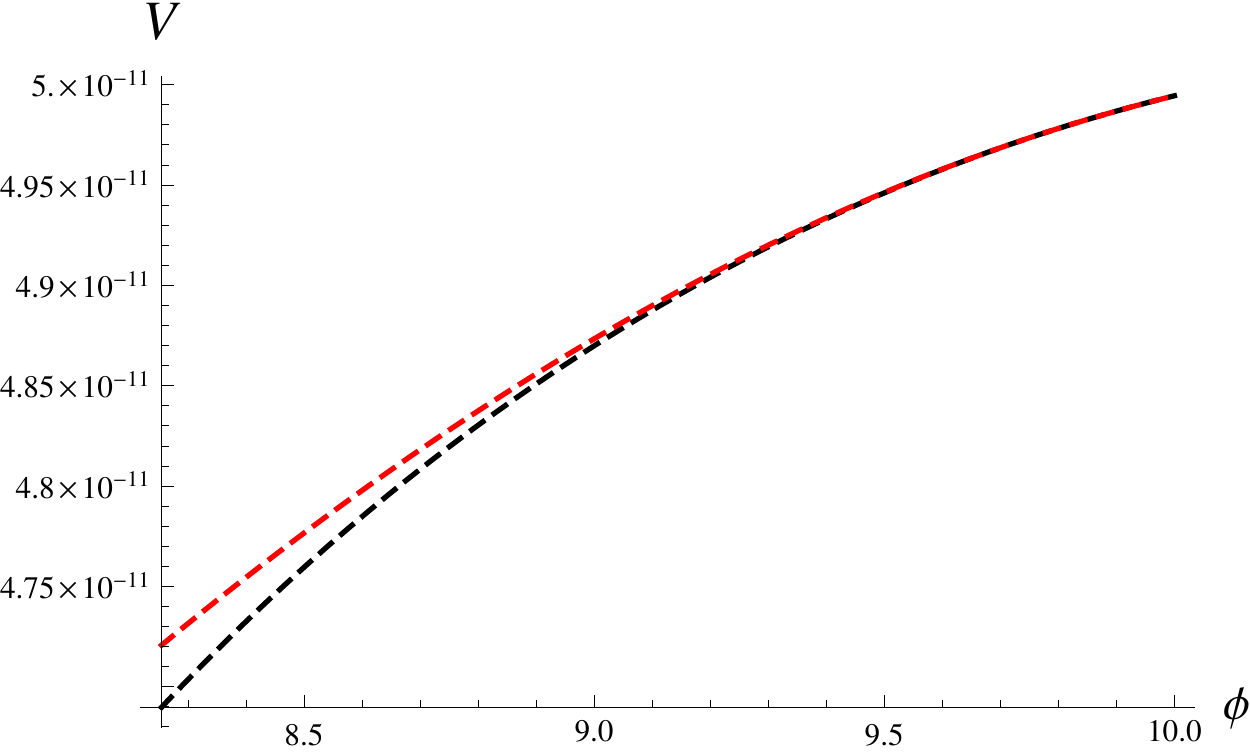}
    \end{minipage}
    \begin{minipage}{.45\textwidth}
    \includegraphics[scale=0.6]{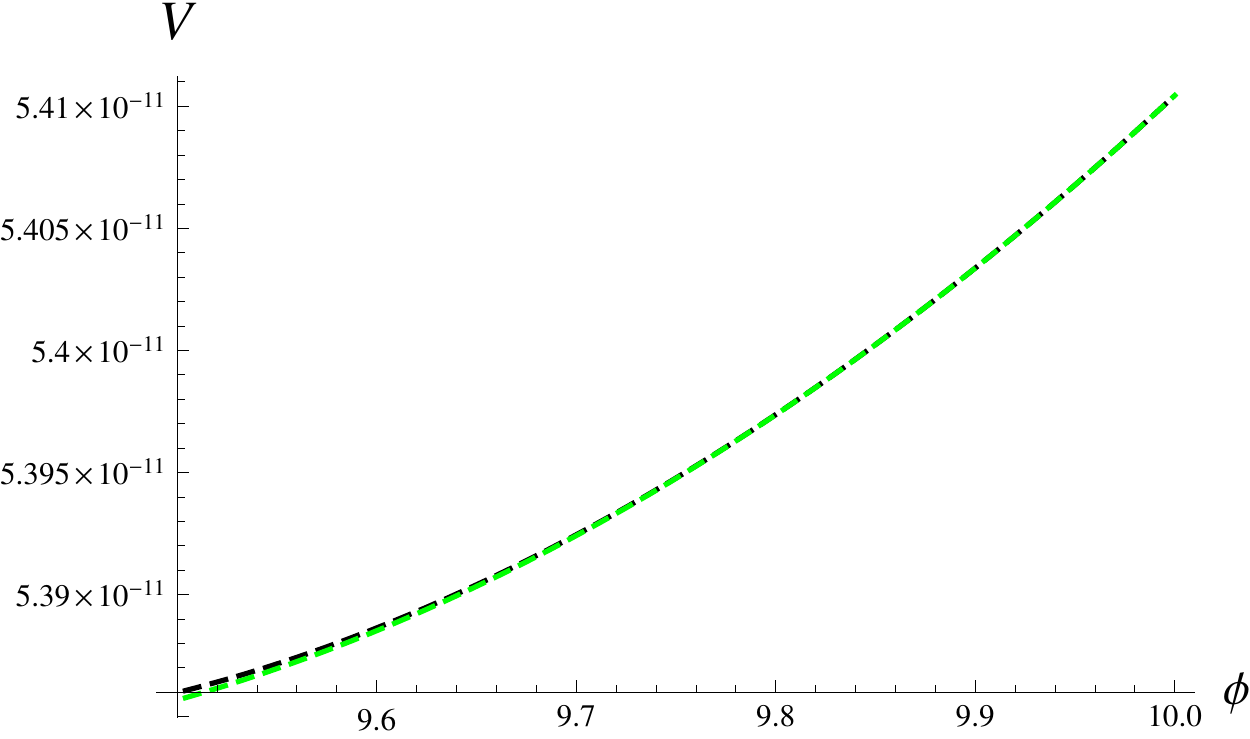}
    \end{minipage}
    \caption{The numerical (black) and analytic (red and green) potentials as functions of $\phi$ are plotted for the same homospectral class. In the plots we use  $H_{2,i}/H_{1,i}= 0.98 \,(1.02)$ in the top (bottom) panel.}
    \label{fig:Vphiplota}
\end{figure}
\begin{figure}
    \includegraphics[scale=0.6]{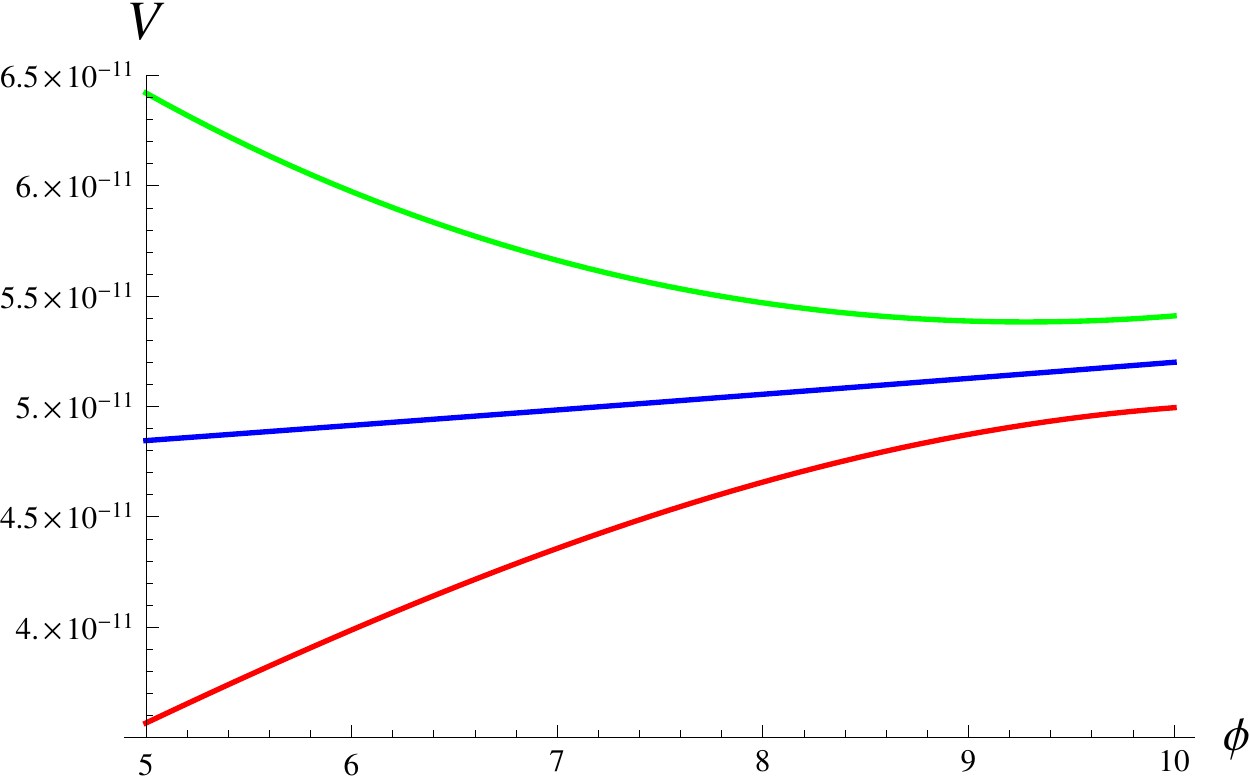}
    \caption{The analytic potential is plotted for different models of the same homospectral class as a function of the field $\phi$. In the plots we use $H_{2,i}/H_{1,i}=1,0.98, 1.02$ for the blue, red, and green lines, respectively.}
    \label{fig:Vphiplotall}
\end{figure}

In Fig.~\ref{fig:Vphiplota} we plot the comparison between the numerically- and analytically-computed homospectral potentials, during an interval of 60 $e$-folds. As can be seen the analytic approximation is in good agreement with the numerical results. In Fig.~\ref{fig:Vphiplotall} we plot the analytically-computed homospectral potentials $V(\phi)$ in \eqs{Va} and \eqn{Vref} as in Fig.~\ref{fig:Vphi},  extending the range of the field beyond the 60 $e$-folds interval.
As explained previously, there is an extra degeneracy for homospectral potentials, related to the choice of $\phi_i$, which we will investigate in a separate work.

\section{Conclusions}

Using purely geometrical arguments it has been shown that there exist classes of homospectral inflationary cosmologies, i.e.\ different expansion histories  producing the same spectrum of comoving curvature perturbations. In this paper we develop a general algorithm to reconstruct the potential of minimally-coupled single scalar fields from an arbitrary expansion history, and we then apply it to homospectral expansion histories to obtain homospectral potentials, providing some numerical and analytical example.

The infinite class of homospectral potentials  depends on two free parameters, the initial energy scale and the initial value of the field, showing that in general it is impossible to reconstruct a unique potential from the curvature spectrum, unless the initial energy scale and the field value are fixed, for instance through observation of primordial gravitational waves.

Our arguments are model independent and can be generalized to any other  theoretical scenarios for which the Sasaki--Mukhanov equation applies, since the origin of the existence of these homospectral models is geometrical, and the evolution of comoving curvature perturbations is completely determined by the expansion history. 

In the future it will be interesting to develop similar reconstruction algorithms for other inflationary models, such as for example Horndeski's theory, but in those cases the presence of a time-varying sound speed will add further degeneracy. This degeneracy could be further increased by the presence of  effective entropy or anisotropy terms in the equation for curvature perturbations \cite{Vallejo-Pena:2019hgv}, which could be studied using the momentum effective sound speed \cite{Romano:2018frb,Romano:2020oov,Rodrguez:2020hot}.

\begin{acknowledgments}
A.G.C. was funded by Agencia Nacional de Investigación y Desarrollo  ANID through the FONDECYT postdoctoral Grant No. 3210512. A.R.L.\ was supported by the Funda\c{c}\~{a}o para a Ci\^encia e a Tecnologia (FCT) through the Investigador FCT Contract No.\ CEECIND/02854/2017 and POPH/FSE (EC), and through the research project PTDC/FIS-AST/0054/2021.
\end{acknowledgments}

\bibliography{Bibliography}
\bibliographystyle{h-physrev4}

\end{document}